\documentclass[11pt]{article}
\usepackage{amsmath}
\usepackage{latexsym}
\usepackage{epsf}
\def\IR{\relax{\rm I\kern-.18em R}}
\def\I1{\relax{\rm 1\kern-.40em 1}}
\def\IZ{\relax{\rm Z\kern-.40em Z}}
\def\be{\begin{equation}}
\def\ee{\end{equation}} 
\begin{document}
\thispagestyle{empty}
\begin{flushright} LPTENS-11/04 \end{flushright}
\vskip 1cm
\begin{center}
{\bf \Large SYMMETRIES AND THE WEAK\\
\vskip 0.3cm
 INTERACTIONS}\\
\vskip 1cm

{\Large JOHN ILIOPOULOS}
\vskip 1cm

Laboratoire de Physique Th\'eorique \\ de L'Ecole Normale Sup\'erieure \\
75231 Paris Cedex 05, France

\vskip 5cm

Talk given at the Cabibbo Memorial Symposium
\vskip 1cm
Rome,  November 12, 2010

\end{center}

\newpage

It is a great honour for me to talk in a Symposium in memory of Nicola Cabibbo
for whom I had great respect and friendship. 

My talk will be a talk about history, given by a non-historian. My purpose is
to trace the origin of 
 the concept which was immediately named by the
high energy physics community {\it the Cabibbo angle}\cite{Cab1}. In doing so, I will occasionally talk about the evolution of other related subjects. Many of these ideas
became part of our common heritage and shaped our understanding of the
fundamental forces of Nature. 

There are many dangers lying in wait for the amateur who attempts to write on
the 
history of science. One is to read the old scientific articles with the light
of today's knowledge, to assume, even sub-consciously, that whatever is clear
now was also clear then. A second is more specific to recent history. 
Because we talk about a period we have witnessed, we tend to trust our memory,
or that of our colleagues. But, as real historians know, and as I have
discovered experimentally, human memory, including one's own, is partial and
selective, especially for events in which one has taken part, even marginally. 
Actors make poor historians, so one should rather try to put his personal 
recollections aside. I do not expect to succeed in producing a work a real
historian would approve, but I hope that
the material I have collected could provide the background
notes he could, eventually, find useful. 
\vskip 0.3 cm

By ``symmetries'' in the weak interactions we mean (i) space-time symmetries,
(ii) global internal symmetries and (iii) gauge symmetries. In all three
fronts the effort to understand their significance has
been one of the most exciting and most rewarding enterprises in modern
physics. It gave rise to the development of novel ideas and concepts whose
importance transcends the domain of weak interactions and encompasses all
fundamental physics. Covering the entire field would be the subject of a book,
so here I will only touch upon a few selected topics which are more directly
related to Nicola's work. I will not talk about the first part, the
establishment of the $V-A$ nature of the weak current and I will not describe
the more modern developments which led to the formulation of the Standard
Model. I will mention some contributions in gauge theories partly because some
of them are not generally known and partly because they touch upon the concept
of universality which is a central theme in my talk. 
\vskip 0.3cm

Although many versions of the history of gauge theories exist already in the
recent literature\cite{Gaugehist}, the message has not yet reached the
textbooks students usually read. I quote a comment from the review by
J.D. Jackson and L.B. Okun: ``... it is amusing how little the authors of
textbooks know about the history of physics.'' Here I  shall just mention some
often forgotten contributions. 

The vector potential
was introduced in classical 
electrodynamics during the first half of the nineteenth century, either implicitly or explicitly, by several authors independently. It appears in some manuscript notes by Carl Friedrich Gauss as early as 1835 and it was fully written by Gustav Kirchoff in 1857, following some earlier work by Franz Neumann and, especially, Wilhelm Weber of 1846. It was soon
noticed that it carried redundant variables and several ``gauge conditions''
were used. The condition, which in modern notation is written as 
$\partial_{\mu}A^{\mu}=0$, was proposed by the Danish mathematical physicist
Ludvig Valentin Lorenz in 1867. Incidentally, most physics books misspell
Lorenz's name as {\it Lorentz}, thus erroneously attributing the condition to
the famous Dutch H.A. Lorentz, of the Lorentz transformations\footnote{In
  French: On ne pr\^ete qu'aux riches.}.  However, for internal symmetries, the concept of
gauge invariance, as we know it to-day, belongs to Quantum Mechanics. It is
the phase of the wave function, or that of the quantum fields, which is not an
observable quantity and produces the internal symmetry
transformations. The local version of these symmetries are the gauge theories
of the Standard Model. The first person who realised that the invariance under
local transformations of the phase of the wave function in the
Schr\"{o}dinger theory implies the introduction of an electromagnetic field
was Vladimir Aleksandrovich Fock in 1926\cite{Fock}, just after
Schr\"{o}dinger wrote his equation. Naturally, one would expect non-Abelian
gauge theories to be constructed following the same principle
immediately after Heisenberg introduced the concept of isospin in 1932. But
here history took a totally unexpected route. 

The development of the General Theory of Relativity offered a new paradigm for
a gauge theory. The fact that it can be written as the theory invariant under
local translations was 
certainly known to Hilbert\cite{Hilbert}. For the next decades it became the
starting point for all studies on theories invariant under local
transformations. The attempt to unify gravitation and electromagnetism via a
five dimensional theory of general relativity is well known under the names of
Theodor Kaluza and Oscar Benjamin Klein\cite{KK}. What is less known is that
the idea was introduced earlier by the Finnish Gunnar Nordstr\"{o}m\cite{Nord}
who had 
constructed a scalar theory of gravitation. In 1914 he wrote a
five-dimensional theory of electromagnetism and showed that, if one assumes
that the fields are independent of the fifth coordinate, the assumption made
later by Kaluza, the electromagnetic vector potential splits into a four
dimensional one and a scalar field identified to his scalar graviton.  
An important contribution from this period is due to Hermann Klaus Hugo
Weyl\cite{Weyl}. He is more known for his 1918 unsuccessful attempt to enlarge
diffeomorphisms to local scale transformations, but, in fact, a byproduct of
this work was a different form of
unification between electromagnetism and gravitation. In his 1929 paper,
 which contains the gauge theory for the Dirac electron, he introduced many concepts
which have become classic, such as the Weyl two-component spinors and the
vierbein and spin-connection formalism. Although the theory is no more scale
invariant, he still used the term {\it gauge invariance}, a term which has
survived ever since.

In particle physics we put the birth of non-Abelian gauge theories in 1954,
with the fundamental paper of Chen Ning Yang and Robert Laurence Mills\cite{YM}. It is
the paper which introduced the $SU(2)$ gauge theory and, although it took some
years before interesting physical theories could be built, it is since that
date that non-Abelian gauge theories became part of high energy physics.  It
is not surprising that they
were immediately named {\it Yang-Mills theories.} The influence of this work in High Energy
Physics has often been emphasised, but here I want to mention some
earlier and little known attempts which, according to 
present views, have followed a quite strange route. 

The first is due to Oscar Klein. In an obscure
conference in 1938 he presented a paper with the title: {\it On the theory of
  charged fields}
\cite{Klein1} in which he attempts to construct an $SU(2)$ gauge theory for
the nuclear forces. This paper is amazing in many ways. First, of course, because
it was done in 1938. He starts from the  discovery of the muon, misinterpreted
as the Yukawa meson, in the old Yukawa theory in which the mesons were assumed
to be vector particles. This provides the physical motivation. The aim is to
write an $SU(2)$ gauge theory unifying electromagnetism and nuclear
forces. Second, and even more amazing, because he follows an
incredibly circuitous road: He
considers General Relativity in a five dimensional space, he compactifies {\it \`a
  la} Kaluza-Klein\footnote{He refers to his 1928 paper but he does not refer
  to Kaluza's 1921 paper. Kaluza is never mentioned. In the course of this
  work I discovered the great interest for the historian of the way people
  cite their own as well as other people's work.}, but he takes the $g_{4\mu}$
components of the metric tensor to be 2x2 matrices. He wants to describe the
$SU(2)$ gauge fields but the matrices he is using, although they depend on
three fields, are not traceless. In spite
of this problem he finds the correct expression for
the field strength tensor of $SU(2)$. In fact, answering an objection by
M{\o}ller, he added a fourth vector field, thus promoting his theory to
$U(1)\times SU(2)$. He added mass terms by hand and it is not clear whether he worried about the resulting breaking of gauge invariance. I cannot find out whether this paper has inspired
anybody else's work because the proceedings of this conference are not
included in the citation index. As far as I know, Klein himself did not follow
up on this idea\footnote{He mentioned this work in a 1956 Conference in
  Berne\cite{Klein2}}. 

The second work in the same spirit is due to Wolfgang
Pauli\cite{Pauli} who in 1953, in a letter to Abraham Pais, as well as in  
 a series of seminars, developed precisely this approach: the construction
of the $SU(2)$ gauge theory as the flat space limit of a compactified higher
dimensional theory of General Relativity. He was closer to the approach
followed to-day because he considered a six dimensional theory with the compact
space forming an $S_2$. He never published this work and I
do not know whether he was aware of Klein's 1938 paper. He had realised that
a mass term for the gauge bosons breaks the invariance\cite{Pauli} and he had an animated argument during
a seminar by Yang in the Institute for Advanced Studies in Princeton in 1954\cite{Y}. What I find surprising is that Klein and Pauli,
fifteen years apart one from the other, decided to construct the $SU(2)$ gauge
theory for strong interactions and both choose to follow this totally  counter-intuitive method.  It
seems that the fascination which General Relativity had exerted on this generation of physicists was such that, for many years, local transformations could not be conceived
independently of general coordinate transformations. Yang
and Mills were the first to understand that the gauge theory of an internal
symmetry takes place in a fixed background space which can be chosen to be
flat, in which case General Relativity plays
no role.

With the work of Yang and Mills gauge theories entered particle physics. 
Although the initial motivation was a theory of the strong interactions, the
first semi-realistic models aimed at describing the weak and electromagnetic
interactions. This story, which led a few years later to the
Standard Model, has been told several times over the last
years\cite{TheRise}, so I shall not follow it up here. I shall only mention a
paper by Sheldon Lee Glashow and Murray Gell-Mann\cite{GG} of 1961 which is
often left out from the history articles. This paper
has two parts: The first extends the Yang-Mills construction, which was
originally done
for $SU(2)$, to arbitrary Lie algebras. The well-known result of associating a
coupling constant to every simple factor in the algebra appeared for the first
time in this paper. Even the seed for a grand unified theory was there. In a
footnote they say: 

``The remarkable universality of the electric charge would be better
understood were the photon not merely a singlet, but a member of a family of
vector mesons comprising a simple partially gauge invariant theory.''

In the second part the authors attempt to apply these
ideas to the strong and weak interactions with interesting implications to the
notion of universality to which I shall come shortly.

\vskip 0.3cm  

By the late fifties the $V-A$ theory was firmly established. The weak current
could be written as a sum of a hadronic and a leptonic part.

\be
\label{weakcur1}
{\cal L}_W= \frac{G_F}{\sqrt{2}}
J^{\mu}(x)J^{\dagger}_{\mu}(x)~~~;~~~J^{\mu}(x)=h^{\mu}(x)+{\ell }^{\mu}(x)
\ee   
where $G_F$ denotes the Fermi coupling constant and $h^{\mu}(x)$ and ${\ell}^{\mu}(x)$ the hadronic and leptonic parts of the weak current. It was easy to guess the form of the leptonic part in terms of the field operators of known leptons. By analogy to the electromagnetic current, we could write:

\be
\label{weakcur2}
{\ell }^{\mu}(x)=\bar {e}(x)\gamma^{\mu} (1+\gamma^5)\nu_{(e)}(x)+...
\ee
where we have used the symbols $e$ and $\nu$ to denote the Dirac spinors of the corresponding particles and the dots stand for the term involving the muon\footnote{The separate identity of the electron and muon neutrinos was not yet established, but it was assumed by some physicists, including Julian Schwinger and Sheldon Glashow.}. There was no corresponding simple form for the hadronic part, since such a form would depend on the knowledge of the dynamics of the strong interactions, in particular the notion of ``elementarity'' of the various hadrons. 

Looking back at the Fermi Lagrangian of equation (\ref{weakcur1}), we see that, since it is a non-renormalisable theory, it can be taken, at best, as an effective theory, in other words only the lowest order terms can be considered. For the leptonic processes this is easy to understand, but for the processes involving $h^{\mu}(x)$ it implies that we should take the matrix elements between eigenstates of the entire strong interaction Hamiltonian. It follows that the statement {\it find the form of $h^{\mu}(x)$} is, in fact, equivalent to the one {\it identify it with a symmetry current of the strong interactions.} As we shall see, this simple fact, which I heard Nicola explaining very clearly in a School in Gif-sur-Yvette\cite{Gif}, was not understood in the early days.

An important step was the hypothesis of the Conserved Vector Current
(C.V.C.)\cite{CVC}. It allowed to identify the strangeness conserving part of
the vector current with the charged components of the isospin
current. Furthermore, it explained the near equality of the coupling constant
measured in muon decay with that of the vector part of nuclear
$\beta$-decay. The non-renormalisation of the latter by the strong
interactions, represented by the pion-nucleon interactions, was correctly
attributed to the conservation of the current. It was the first concrete
realisation of the concept of universality, which, at this stage, was taken to
mean ``equal couplings for all processes''. However, the connection with the
algebraic properties of the currents came a bit later. 

With the introduction of strange particles the picture became more
complicated. Several schemes were proposed to extend isospin to a symmetry
including the strange particles, which I will not review here. I will
concentrate on the evolution of the ideas referring to the weak interactions.

The first contribution I want to mention comes from the young CERN Theory
Group which was established in Geneva in 1954 and it is the work of Bernard
d'Espagnat and Jacques Prentki. They had already worked in
various higher symmetry schemes and in 1958 they addressed the question of the
weak interactions\cite{Des-Pr1}. The title of the paper is {\it A tentative
  general scheme for weak interactions} and, for the first time, a
comprehensive picture of the whole hierarchy of symmetries for all
interactions is clearly presented. They were working under the assumption of
$O(4)$ being the higher symmetry group. Four levels were considered: (i) The
{\it very strong interactions} are invariant under $O(4)$. (ii) The {\it
  medium strong interactions} break $O(4)$ and leave one $SU(2)$ (isospin) and
the third component of the other (strangeness), invariant. (iii) The {\it
  electromagnetic interactions} conserve only the third components of the two
$SU(2)$'s (electric charge and strangeness). (iv) Finally, the {\it weak
  interactions} which, like the medium strong ones, conserve an $SU(2)$, but
which is a different subgroup of $O(4)$. Thus, strangeness violation is
presented as the result of a mismatch between the medium strong and the weak interactions. Let me add here that the idea of introducing medium strong interactions was already known, but in its early versions it was supposed to describe the interactions of $K$ mesons as opposed to those of pions which were the very strong ones. The correct scheme, as we know it to-day, appeared for the first time in d'Espagnat and Prentki's paper. In reading this very lucid and beautiful paper one may not understand why these authors failed to discover the Cabibbo theory immediately after the introduction of $SU(3)$. This is an example of the first danger for the historian I mentioned in the introduction, namely reading old papers with to-day's knowledge. As we shall see shortly, matters were not that simple.

I skip a couple of other contributions which should be included in a complete
history article and I come to a very important paper by M. Gell-Mann and
Maurice L\'evy\cite{GL} with the title {\it The Axial Vector Current in Beta
  Decay}. It presented the well-known {\it $\sigma$-model}, both the linear
and non-linear versions, which became the paradigm for chiral symmetry. In the
Introduction section they note that the radiative corrections to the
$\mu$-decay amplitude which had just been computed gave a slight discrepancy
with C.V.C., namely $G_V/G_{\mu}=0.97\pm 0.01$. In the published version at
this point there is a {\it Note added in proof}. I copy: 

``Should this discrepancy be real ({\it $g_V \neq 1$})
it would probably indicate a total or partial failure of the conserved vector
current idea. It might also mean, however, that the current is conserved but
with $g_V <1$. Such a situation is consistent with universality if we consider
the vector current for $\Delta S =0$ and $\Delta S =1$ together to be
something like:

\begin{equation}
GV_{\alpha}+GV_{\alpha}^{(\Delta
  S=1)}=G_{\mu}\bar{p}\gamma_{\alpha}(n+\epsilon
\Lambda)(1+\epsilon^2)^{-1/2}+...\nonumber
\end{equation}
and likewise for the axial vector current. If $(1+\epsilon^2)^{-1/2}$ =0.97,
then $\epsilon^2$=.06, which is of the right order of magnitude for explaining
the low rate of $\beta$ decay of the $\Lambda$ particle. {\it There is, of
  course, a renormalization factor for that decay, so we cannot be sure that
  the low rate really fits in with such a picture.}'' (my italics). 

We see that the idea of considering a linear combination of the strangeness conserving and strangeness changing currents is there with the correct order of magnitude for the coefficients, but this is presented as a coincidence which was expected to be spoiled by uncontrollable renormalisation effects. Let me notice here that the paper was meant to be a model for the $\Delta S$=0 axial vector current and the properties of the $\Delta S$=1 vector current were just a side remark in a footnote.   

Continuing in chronological order, I come to the 1961 paper by Glashow and
Gell-Mann\cite{GG}.  After looking at gauge theories for higher groups I
mentioned above, the authors try to apply the non-Abelian gauge theories to
particle physics. They study both strong interactions, for which they attempt
to identify the gauge bosons with the vector resonances which had just been
discovered, as well as weak interactions. The currents were written in the
Sakata model\cite{Sak}, although no reference to Sakata is given. Notice also
that Gell-Mann had just written the paper on {\it The eightfold
    way}\cite{SU3}, but here they do not want to commit themselves on $SU(3)$ as the
symmetry group of strong interactions, so they do not exploit the property of
the currents to belong to an octet. The paper is remarkable in many
aspects, besides the ones I mentioned already in extending Yang-Mills to
higher groups. For the weak interactions it considers the Glashow $SU(2)\times
U(1)$ model\cite{Gl} and it correctly identifies the problems
related to the absence of strangeness changing neutral currents and the small
value of the $K^0_1-K^0_2$ mass difference. The question of universality is
addressed in a footnote (remember that they were working in the Sakata model):

``Observe that the sum of the squares of the coupling strengths to
strangeness-saving charged currents and to strangeness-changing charged currents is just
the square of the universal coupling strength. Should the gauge principle be
extended to leptons - at least for the charged currents - the equality between
$G_V$ and $G_{\mu}$ is no longer the proper statement of universality, for in
this theory $G_V^2+G_{\Lambda}^2=G_{\mu}^2$ ($G_{\Lambda}$ is the {\it
  unrenormalized} (their italics) coupling strength for $\beta$-decay of
$\Lambda$)''. 

I do not know why this paper has not received the attention it deserves, but
this is partly due to the authors themselves, especially Gell-Mann, who rarely
referred to it\footnote{There is a reference in the G.I.M. paper\cite{GIM}
  where the related problems were solved.}. 

As I said above, it was the time
 $SU(3)$ was introduced\cite{SU3}, so, it was natural to apply it to weak
interactions. I want first to present an attempt by d'Espagnat and
Prentki\cite{Des-Pr2} who followed the lines of their 1958 paper. They
reconsider their former $O(4)$ theory and tried to adapt it to $SU(3)$. In the
Introduction they make their assumptions explicit and write:

``Before we begin, it is proper to insist on how uncertain and speculative
such attempts necessarily are: a) it is not at all proved that the strong
interactions really have anything to do with $SU(3)$, b) {\it even if they
  have, the statement that the same is true, in some way, for weak
  interactions is just a guess}....'' (my italics).

Of course, to-day such a statement sounds strange, but we must remember we are
in 1962. The simple fact we explained above, namely that in the effective
Fermi theory the hadronic weak current is an operator acting in the space of
hadrons, {\it i.e.} in the space of eigenstates of the strong interaction
Hamiltonian, was not fully understood. d'Espagnat and Prentki had not realised
that their two assumptions were not independent. In spite of that and taking
into account their beautiful paper of 1958, I would expect them to go ahead
and assign $SU(3)$ transformation properties to the weak current. In fact they
start this way and in section 3 of their paper we can find the correct form of
$h_{\mu}(x)$  as a superposition of a $\Delta S$=0 and a $\Delta S$=1 part with
an angle they call $\alpha$. Then they proceed to show that in a
current x current theory the two empirical selection rules $|\Delta S| \leq 1$
and $|\Delta I| <3/2$ for non-leptonic processes are related. And they stop
there! They do not look at all at the semi-leptonic processes. In their paper
leptons are mentioned only at the last paragraph. It seems that they were
misled by an erroneous experiment claiming evidence for $\Delta S =-\Delta Q$
decays. Indeed, there was a single event $\Sigma^+\rightarrow \mu^++n+\nu$ reported in an emulsion experiment\cite{WrExp} which dates from that period, but I do not know whether it is the right one, since they do
not mention it anywhere in the paper. 

The following year a short paper appeared as a CERN preprint\cite{Cab1}. The
author was a young visitor from Italy, Nicola Cabibbo. He took over the idea
of a current which forms an angle with respect to medium strong interactions,
but he carried it to its logical conclusion. This form allows for the only
consistent definition of universality. Using modern quark language his remark
can be translated into the statement that, with one quark of charge 2/3 and two
quarks of charge -1/3, one could always construct one hadronic current which is
coupled to leptons and one which is not. He naturally defined universality by
the assumption that the coupled current involves the same coupling constant as
the purely leptonic processes. As Glashow and Gell-Mann, he assigned the
current to an octet of $SU(3)$, but he was in the eightfold way scheme
and not in the Sakata model. This allowed him to compare strangeness
conserving and strangeness changing semi-leptonic decays and show that the
scheme agreed with experiment, thus putting the final stone into the $SU(3)$
edifice\cite{Velt}. The name which was attached to this paper was {\it The Cabibbo angle}, but in
fact, the most important point was the proof that the hadronic weak current
has the right transformation properties under $SU(3)$. I do not know whether
he was unaware of the wrong experimental result or whether he showed the good 
physical judgement to ignore it, but he clearly understood all the underlying
physics. Since it appeared as a CERN preprint, I included it in my 1996
article on the Physics in the CERN Theory Division\cite{Ilio}, where I wrote:

``There are very few articles in the scientific literature in which one does
not feel the need to change a single word and Cabibbo's is definitely one of
them. With this work he established himself as one of the leading theorists in
the domain of weak interactions.''

The total number of citations is not a reliable criterion for the importance
of a scientific article, but, not surprisingly, Cabibbo's paper is among the
most cited ones in high energy physics. Concerning citations, I found it
interesting to see how each article was cited from the protagonists in this
field. Just two examples: Gell-Mann, in his 1964 article on current
algebras\cite{CA} refers to himself and Cabibbo. At the same year d'Espagnat
gave a very beautiful set of lectures which were published as a CERN
report\cite{Desp}. The title was {\it $SU(3)$ et Interactions Faibles.} He
cites Feynman and Gell-Mann for CVC and Cabibbo\footnote{It is remarkable that
  he does not cite any of his own papers.}.

\vskip 0.3cm

Cabibbo's scientific work spans five decades, until the very last days of his
illness. His name will remain in the Physics text books and he will continue
to inspire the young physicists. But those of us who had the good fortune to
know him will miss his sound judgement, his enthusiasm for physics, but also
his gentle and friendly manners. He would never get angry and shout,
only his polite smile would occasionally show disapproval. 

Cabibbo will be always with us, but we shall miss Nicola.

\end{document}